\documentclass[aps,pra,twocolumn]{revtex4}
\usepackage{graphicx}
\usepackage{dcolumn}
\usepackage{amsmath}
\usepackage{amssymb}
\usepackage{float}
\usepackage{epstopdf}
\def\erfc{{\rm erfc}}
\def\erf{{\rm erf}}
\def\rv{{\bf r}}
\def\sv{{\bf s}}
\def\yv{{\bf y}}
\def\xv{{\bf x}}
\def\Rv{{\bf R}}
\def\kv{{\bf k}}
\def\qv{{\bf q}}

\def\ec{\epsilon_c}

\def\beq{\begin{equation}}
\def\eeq{\end{equation}}

\begin{document}
\title{Properties of short-range and long-range correlation energy density functionals
from electron-electron coalescence}
\author{Paola Gori-Giorgi and Andreas Savin}
\affiliation{Laboratoire de Chimie Th\'eorique, CNRS,
Universit\'e Pierre et Marie Curie, 4 Place Jussieu,
F-75252 Paris, France}
\date{\today}

\begin{abstract}
The combination of density functional theory with other approaches
to the many-electron problem through the
separation of the electron-electron interaction into a short-range
and a long-range contribution is a promising method, which is
raising more and more interest in recent years. In this work some
 properties of the corresponding correlation energy
functionals are derived by studying
the electron-electron coalescence condition for a modified (long-range-only)
interaction. A general relation for 
the on-top (zero electron-electron distance) pair density is derived, 
and its usefulness is
discussed with some examples. For the special case of the uniform electron gas, 
a simple parameterization of the on-top pair density for a long-range only 
interaction is presented and supported by calculations within the
``extended Overhauser model''. The results of this work can be used to
build self-interaction corrected short-range correlation energy functionals.
\end{abstract}

\maketitle
\section{Introduction}
In recent years, there has been a growing interest in approaches that combine
density functional theory~\cite{kohnnobel,science,FNM} (DFT) with other 
approximate methods to treat the many-electron problem. 
In most cases, this combination is achieved by splitting
the Coulomb electron-electron interaction $1/r_{12}$ into a short-range (SR) and a 
long-range (LR) part (see, e.g.,
Refs~\onlinecite{kohn_mm,ikura,scuseria,vaffa,erf,sav_madrid,julien,janos,yamaguci}). 
The idea is to use different, appropriate approximations for the long-range and
the short-range contributions to the exchange and/or correlation energy
density functionals of the Kohn-Sham (KS) scheme, to treat, e.g., near-degeneracy
effects or van der Waals forces. These approaches are often inspired by
the observation that long-range correlations are not
well treated by local or semilocal density functionals, but can be dealt with
by other techniques, like the standard wavefunction methods of quantum chemistry. 
Conversely, correlation effects due to the short-range part of the 
electron-electron interaction can be well described by local or semilocal 
functionals (appropriately modified).
\par
\begin{figure}
\includegraphics[width=7.2cm]{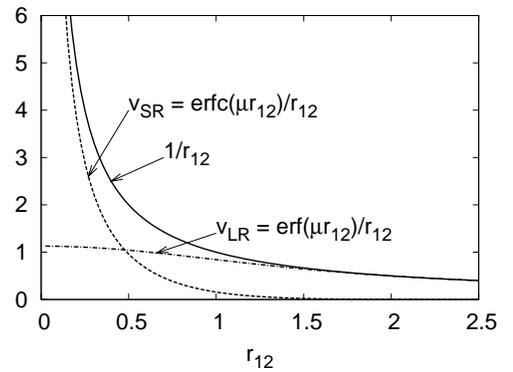} 
\caption{The splitting of the Coulomb interaction $1/r_{12}$ into
a short-range (SR) and a long-range (LR) part as defined in
 Eq.~(\ref{eq_split}). Here $\mu=1$. When $\mu\to\infty$ we have
$v_{\rm SR}\to 0$ and $v_{\rm LR}\to 1/r_{12}$, and when $\mu\to 0$
we have $v_{\rm SR}\to 1/r_{12}$ and $v_{\rm LR}\to 0$. }
\label{fig_esempio}
\end{figure}
The error function and its complement (see Fig.~\ref{fig_esempio}),
\beq
\frac{1}{r_{12}}=v_{\rm SR}^\mu(r_{12})+v_{\rm LR}^\mu(r_{12})=
\frac{\erfc(\mu r_{12})}
{r_{12}}+\frac{\erf(\mu r_{12})}{r_{12}},
\label{eq_split}
\eeq
are often used~\cite{ikura,erf,scuseria,julien,janos} 
for the splitting of the Coulomb interaction, 
since they yield analytic matrix elements for both Gaussians 
and plane waves, i.e., the most common basis functions in quantum chemistry 
and solid state physics, respectively. The parameter $\mu$ controls the range 
of the decomposition. Correspondingly, the universal Coulombic functional of
the electron density $n(\rv)$, 
$F[n]$, as defined in the constrained search formalism~\cite{mel},
\beq
F[n]=\min_{\Psi\to n}\langle \Psi|T+V_{ee}|\Psi\rangle
\eeq
can be divided into a long-range part and a complementary
short-range part, $F[n]  =  F_{\rm LR}^\mu[n]+\overline{F}_{\rm SR}^\mu[n]$,
\begin{eqnarray}
F_{\rm LR}^\mu[n] & = & \min_{\Psi^\mu\to n}\langle \Psi^\mu|T+V_{\rm LR}^\mu|\Psi^\mu\rangle 
\nonumber \\
\overline{F}_{\rm SR}^\mu[n] & = & F[n]-F_{\rm LR}^\mu[n],
\label{eq_LR-SR}
\end{eqnarray}
or, alternatively, into a short-range part and a complementary long-range
part
\begin{eqnarray}
F_{\rm SR}^\mu[n] & = & \min_{\tilde{\Psi}^\mu\to n}\langle 
\tilde{\Psi}^\mu|T+V_{\rm SR}^\mu|\tilde{\Psi}^\mu\rangle 
\nonumber \\
\overline{F}_{\rm LR}^\mu[n] & = & F[n]-F_{\rm SR}^\mu[n].
\label{eq_SR-LR}
\end{eqnarray}
These two decompositions lead to different exchange-correlation energy 
functionals that need to be approximated; they are compared in 
Ref.~\onlinecite{jul-sav}, where their advantages and disadvantages are discussed.

In the present work we focus on the properties of the long-range and short-range
correlation functionals that come from the modification of the electron-electron 
interaction at short distances, i.e. those properties that are due to the change
in the electron-electron coalescence conditions. This means that we are only
concerned with the functionals of the decomposition 
of Eq.~(\ref{eq_LR-SR}), which involve a many-body wavefunction $\Psi^\mu$ 
of a system with an electron-electron interaction that is softer than
$1/r_{12}$ for small $r_{12}$ (see Fig.~\ref{fig_esempio}). 
This decomposition is the one used in the approaches of 
Refs.~\onlinecite{ikura,vaffa,erf,sav_madrid,julien,janos}.

The paper is organized as follows. In Sec.~\ref{sec_defs} we define the 
long-range and short-range correlation
energy functionals that are the object of the present study. 
In Sec.~\ref{sec_cusp} we
analyze the short-range properties of the pair density of a general
many-electron system
with interaction $\erf(\mu r_{12})/r_{12}$ when $\mu$ gets larger and larger: we
derive an expansion for $\mu\to\infty$ of the on-top (zero electron-electron
distance) pair density, and,
following (and partly correcting) the work 
of Ref.~\onlinecite{julien}, 
an expansion for $\mu\to\infty$ of the
correlation energy functionals. Section~\ref{sec_elegas} is devoted to the special
case of the uniform electron gas: starting from the
exact high-density limit, a simple parameterization
of the on-top pair density as a function of $\mu$ is proposed, and is 
favorably compared
with the results obtained from the 
``extended Overhauser model'' \cite{Ov,GP1} for the same
quantity. The last Sec.~\ref{sec_approx} explains, with some examples, how the
results of this work can be used to build self-interaction corrected
 approximations for short-range correlation functionals.
\par
Hartree atomic units are used throughout this work.

\section{Definitions and basic equations}
\label{sec_defs}
When the universal functional $F[n]$ is decomposed as in Eq.~(\ref{eq_LR-SR}),
we have a model system, whose wavefunction is denoted $\Psi^\mu$, which has
the same density $n(\rv)$ of the physical system and electron-electron
interaction $\erf(\mu r_{12})/r_{12}$. When $\mu\to 0$ this model system
becomes the Kohn-Sham system, with no electron-electron interaction, while
when $\mu\to\infty$ the model system approaches the physical 
one, with interaction
$1/r_{12}$. By definition, the density is the same for all values of $\mu$.
In the approach of Refs.~\onlinecite{vaffa,erf,sav_madrid,julien}
the model system at a fixed $\mu$ 
is treated with a multideterminantal 
wavefunction. In general, if $\mu$ is not too large, 
few determinants describe $\Psi^\mu$ quite
accurately (because of the smaller interaction, and also because of the
 absence of the electron-electron cusp);
the larger is the chosen value of $\mu$, the larger is the needed configuration
space and thus the computational
cost. The remaining
part of the energy is provided by the complementary functional 
$\overline{F}_{\rm SR}^\mu[n]=F[n]-F_{\rm LR}^\mu[n]$ 
of Eq.~(\ref{eq_LR-SR}), which can be divided
into Hartree, exchange, and correlation contributions in the usual way
(for an alternative separation of exchange and correlation, see 
Ref.~\onlinecite{tca}),
\begin{eqnarray}
\overline{E}_{\rm H,\, SR}^\mu[n] & = &
 \frac{1}{2}\int d\rv \int d\rv' n(\rv)n(\rv')v_{\rm SR}^\mu(|\rv-\rv'|) \\
\overline{E}_{x,\, {\rm SR}}^\mu[n] & = & 
\langle \Phi |V_{\rm SR}^\mu|\Phi\rangle-\overline{E}_{\rm H,\, SR}^\mu[n] \\
\overline{E}_{c,\, {\rm SR}}^\mu[n]& = & \overline{F}_{\rm SR}^\mu[n]-
\overline{E}_{\rm H,\, SR}^\mu[n]-\overline{E}_{x,\, {\rm SR}}^\mu[n],
\end{eqnarray}
where $\Phi$ is the Kohn-Sham determinant. Notice that the Hartree and the
exchange functional are linear in the interaction, so that 
$\overline{E}_{\rm H,\, SR}^\mu[n]  =  E_{\rm H,\, SR}^\mu[n]$ and
$\overline{E}_{x,\, {\rm SR}}^\mu[n]  =  E_{x,\, {\rm SR}}^\mu[n]$,
where $E_{\rm H,\, SR}^\mu[n]$ and  $E_{x,\, {\rm SR}}^\mu[n]$ are the short-range
functionals of the decomposition of Eq.~(\ref{eq_SR-LR}). The correlation
energy, instead, depends on the wavefunction $\Psi^\mu$ and we thus
have $\overline{E}_{c,\, {\rm SR}}^\mu[n]\neq E_{c,\, {\rm SR}}^\mu[n]$. The
complementary correlation functional $\overline{E}_{c,\, {\rm SR}}^\mu[n]$ is
 the difference between the usual Coulombic correlation energy
$E_c[n]$, and the long-range correlation energy functional 
$E_{c,\, {\rm LR}}^\mu[n]$,
\begin{eqnarray}
E_{c,\, {\rm LR}}^\mu[n]& = & \langle\Psi^\mu|T+V_{\rm LR}^\mu|\Psi^\mu\rangle
-\langle\Phi|T+V_{\rm LR}^\mu|\Phi\rangle \\
\overline{E}_{c,\, {\rm SR}}^\mu[n] & = & E_c[n]-E_{c,\, {\rm LR}}^\mu[n].
\end{eqnarray}
In what follows we study the short-range functional 
$\overline{E}_{c,\, {\rm SR}}^\mu[n]$ or, equivalently the long-range
functional $E_{c,\, {\rm LR}}^\mu[n]$, in the limit of large $\mu$, i.e.,
when the model system described by $\Psi^\mu$ is approaching the physical
system. 

Following Toulouse {\it et al.}~\cite{julien}, 
we start from the Helmann-Feynmann theorem which gives 
\beq
\frac{\partial}{\partial \mu}\overline{E}_{c,\, {\rm SR}}^\mu[n]=
-\frac{2}{\sqrt{\pi}}\int_0^\infty 4\pi r_{12}^2\, f_c^\mu(r_{12})\,
e^{-\mu^2 r_{12}^2}\, d r_{12},
\label{eq_hel}
\eeq
where the spherically and system-averaged pair density (or intracule density)
$f^\mu(r_{12})$ is obtained by integrating $|\Psi^\mu|^2$ over all variables but
$r_{12}=|\rv_2-\rv_1|$: we first define the spherical average of the
pair density (with $\rv=\rv_1$),
\begin{eqnarray}
\tilde{P}_2^\mu(\rv,r_{12}) = \frac{N(N-1)}{2}\sum_{\sigma_1...\sigma_N} \times \nonumber \\ 
 \int |\Psi^\mu(\rv,\rv_{12},\rv_3,...,\rv_N)|^2
\frac{d\Omega_{\rv_{12}}}{4\pi} d\rv_3...d\rv_N,
\label{eq_p2sph}
\end{eqnarray}
and then integrate over all reference positions $\rv$,
\begin{equation}
f^\mu(r_{12}) = \int \tilde{P}_2^\mu(\rv,r_{12})\,d\rv .
\label{eq_intra}
\end{equation}
 The correlated part of $f^\mu(r_{12})$ appearing in
Eq.~(\ref{eq_hel}) is  $f_c^\mu=f^\mu-f_{\rm KS}$, where 
$f_{\rm KS}(r_{12})$ is obtained by  replacing $\Psi^\mu$ 
with  the Kohn-Sham determinant $\Phi$ in Eq.~(\ref{eq_p2sph}).
\par
The correlated intracule $f_c^\mu(r_{12})$ can be expanded in its Taylor series 
around $r_{12}=0$ up to some order $M$,
\beq
f_c^\mu(r_{12})=\sum_{n=0}^{M-1} c_n(\mu)\ r_{12}^n+O(r_{12}^M).
\eeq
By inserting this expansion into Eq.~(\ref{eq_hel}) we find~\cite{julien}
\beq
\frac{\partial}{\partial \mu}\overline{E}_{c,\, {\rm SR}}^\mu[n]=
-4 \sqrt{\pi}\sum_{n=0}^{M-1}\frac{c_n(\mu)}{\mu^{n+3}}\,
\Gamma\left(\frac{n+3}{2}\right)
+O\left(\frac{1}{\mu^{M+3}}\right).
\label{eq_dEc-largemu}
\eeq
This means that when $\mu\to\infty$, i.e., when we are approaching the
full interaction $1/r_{12}$, the correlation energy functional 
$\overline{E}_{c,\, {\rm SR}}^\mu[n]$ has an expansion in powers of
$\mu^{-1}$  whose coefficients are determined by the short-range
behavior of $f_c^\mu$. In order to determine as many coefficients as
possible in the expansion of Eq.~(\ref{eq_dEc-largemu}) we thus have
to know how the $c_n(\mu)$ behave in the limit of large $\mu$.
In Ref.~\onlinecite{julien} the same expansion was considered, and the first
two terms were obtained by simply inserting in Eq.~(\ref{eq_dEc-largemu}) 
the values of $c_0$ and $c_1$ for the physical system (with Coulomb interaction).
In the next Sec.~\ref{sec_cusp} we study the general problem of determining
the short-range behavior of $f^\mu(r_{12})$ in the limit $\mu\to\infty$. We find
the same result of Ref.~\onlinecite{julien} for the first term ($\propto \mu^{-3}$)
of Eq.~(\ref{eq_dEc-largemu}), but a different result for the second 
term ($\propto \mu^{-4}$), and we
explain why. Moreover, we obtain the first-order (in $\mu^{-1}$) term of the
expansion for large $\mu$ of the on-top $f^\mu(0)$.

\section{Short-range behavior of a system with interaction
$\erf(\mu r_{12})/r_{12}$ when $\mu\to\infty$}
\label{sec_cusp}
We start from the Schr\"odinger equation for the wavefunction $\Psi^\mu$,
\begin{eqnarray}
& & H^\mu\,\Psi^\mu  =  E^\mu\,\Psi^\mu,  \label{eq_Hmu}\\
& & H^\mu  =  -\sum_{i=1}^N\frac{\nabla^2_i}{2}+\sum_{i> j=1}^N
\frac{\erf(\mu|\rv_i-\rv_j|)}{|\rv_i-\rv_j|}+\sum_{i=1}^N v_\mu(\rv_i),
\nonumber 
\end{eqnarray}
where the one-body potential $v_\mu(\rv)$ keeps the density equal to
the one of the physical system for every $\mu$.
\par
When $\mu\to\infty$  the interaction $\erf(\mu\,r_{12})/r_{12}$
gets larger and larger for small $r_{12}$ ($r_{12}\ll 1/\mu$). If we thus fix a finite
but very large value of $\mu$, we can use the same arguments that
lead to the derivation of the electron-electron cusp condition for the Coulomb 
interaction~\cite{kato,kimball,kutzel,ugalde}, i.e., we can isolate two coalescing
electrons (say, 1 and 2) in the hamiltonian of Eq.~(\ref{eq_Hmu}), and switch to 
variables $\rv_{12}=\rv_1-\rv_2$ and $\Rv=(\rv_1+\rv_2)/2$.
In the limit $r_{12}=|\rv_{12}|\to 0$ there must be a term in $H^\mu \Psi^\mu$
that compensates the divergence (or, more precisely, the very large value) of
$\erf(\mu\,r_{12})/r_{12}$. As for the Coulombic systems,
this compensation comes from the relative kinetic energy term, and to
determine the small $r_{12}$ behavior of the spherical average of 
$|\Psi^\mu|^2$ we only need to look
at the Schr\"odinger equation for the relative motion of 
two electrons~\cite{kato,kimball,kutzel}
 approaching each other with relative angular momentum $\ell=0$. 
Higher $\ell$, in fact, will contribute to $|\Psi^\mu|^2$ to
orders $r_{12}^{2\ell}$ in the limit of small $r_{12}$. Only in the case of a fully
polarized system the case $\ell=1$ must be 
considered to determine the $r_{12}\to 0$ behavior of $|\Psi^\mu|^2$, since
only odd $\ell$ are allowed~\cite{kimball} (triplet symmetry); this case
is considered in Appendix~\ref{app_l}. The rare case of unnatural parity 
singlet states~\cite{kutzel}
 (which needs $\ell=2$) is not considered in this work.

As it was done for the Coulomb electron-electron 
interaction~\cite{kato,kimball,kutzel},
we thus focus on the relative wavefunction $\psi^\mu(r_{12})$ for two electrons in 
the $\ell=0$ state. By defining $x=r_{12}$ and $u^\mu(x)=x\,\psi^\mu(x)$, the
relevant Schr\"odinger-like equation reads
\beq
\left[-\frac{d^2}{dx^2}+\frac{\erf(\mu x)}{x}\right]u^\mu(x)=\mathcal{E}^\mu(\xv,\Rv,
\rv_3,...,\rv_N)u^\mu(x),
\label{eq_relative}
\eeq
where $\mathcal{E}^\mu$ is a complicated operator that does not affect the result
as long as it remains bounded when $\mu\to\infty$ and $x\to 0$, as it is reasonable to
assume from the hamiltonian of Eq.~(\ref{eq_Hmu})~\cite{kato,kimball,kutzel}.
We change variable $y=\mu x$, and divide both members of 
Eq.~(\ref{eq_relative}) by $\mu^2$ to obtain
\beq
\left[-\frac{d^2}{dy^2}+\frac{1}{\mu}\frac{\erf(y)}{y}\right]u^\mu(y)=
\frac{1}{\mu^2}\mathcal{E}^\mu(\yv,\Rv,
\rv_3,...,\rv_N)u^\mu(y).
\label{eq_relativey}
\eeq 
We then expand $u^\mu(y)$ for large $\mu$, 
\beq
u^\mu(y)=u^{(\infty)}(y)+\tfrac{1}{\mu}u^{(-1)}(y)+
O\left(\tfrac{1}{\mu^2}\right), 
\label{eq_expau}
\eeq
insert this expansion into Eq.~(\ref{eq_relativey}), and impose
 that the left-hand-side be of order $\mu^{-2}$ as the right-hand side. 
With the boundary condition that $\psi^\mu(x)$ is finite at $x=0$, we obtain
\begin{eqnarray}
u^{(\infty)}(y) & = & a y \label{eq_u0} \\
\frac{d^2u^{(-1)}(y)}{dy^2} & = & a\,\erf(y), \label{eq_u1}
\end{eqnarray}
and we find that
the final solution for $\psi^\mu(x)$ from Eqs.~(\ref{eq_u0})-(\ref{eq_u1}) is
\beq
\psi^\mu(x)  = 
a\left[1+x\,p_1(\mu x)+\frac{1}{\sqrt{\pi}\mu}+\frac{A_1}{\mu}+...\right],
\eeq
where $A_1$ is a constant coming from the integration of
 Eq.~(\ref{eq_u1}) that is not determined by the condition
that $\psi^\mu(x)$ is finite in $x=0$, and $a$ determines the value
$\psi(0)$ for the Coulombic system ($\mu=\infty$). The function $p_1(y)$
is given by
\beq
p_1(y) =  \frac{e^{-y^2}-2}{2\sqrt{\pi}\, y}
+\left(\frac{1}{2}+\frac{1}{4\,y^2}\right) \erf(y),
\label{eq_p1}
\eeq
and has the following asymptotic behaviors
\begin{eqnarray}
p_1(y\to 0) & = & \frac{y}{3\sqrt{\pi}}+O(y^3) \label{eq_smallp1} \\
p_1(y\to\infty) & = & \frac{1}{2}-\frac{1}{\sqrt{\pi}\,y}+O\left(\frac{1}{y^2}\right).
\label{eq_expap1}
\end{eqnarray}
The spherically- and system-averaged pair-density of Eq.~(\ref{eq_intra}) has
thus, to leading orders in $1/\mu$ for large $\mu$, the small-$r_{12}$ expansion
\beq
f^\mu(r_{12})=f(0)\left[1+2\,r_{12}\, p_1(\mu r_{12})+\frac{2}{\sqrt{\pi}\mu}+
\frac{2 A_1}{\mu}\right],
\label{eq_expaf}
\eeq
where $f(0)$ is the on-top value corresponding to the full interacting system
[$f(0)$ is proportional to $a^2$, where $a$ determines $u^{(\infty)}(y)$ 
 in Eq.~(\ref{eq_u0})].
Equation~(\ref{eq_expap1}) tells us that
if in Eq.~(\ref{eq_expaf}) we fix $r_{12}$ equal to a 
small value $r_0\ll 1$, and then
let $\mu$ go to $\infty$ we recover the Coulombic cusp, 
$f(r_0)=f(0)(1+r_0+...)$. But for any finite large $\mu$, 
Eq.~(\ref{eq_smallp1}) shows that we always
obtain a quadratic behavior for small $r_{12}$, 
$f^\mu (r_{12})=f(0)(1+2 r_{12}^2\mu/3\sqrt{\pi}+...)$.
This is how the cuspless wavefunction corresponding to the interaction
$\erf(\mu r_{12})/r_{12}$ develops the Coulombic cusp in the $\mu\to\infty$
limit. An alternative derivation of Eq.~(\ref{eq_expaf}), more similar
to what one usually does for the Coulombic 
cusp~\cite{kato,kimball,kutzel,ugalde}, is reported in 
Appendix~\ref{app_2balle}.
\par
To obtain the complementary short-range correlation functional we can
insert Eq.~(\ref{eq_expaf}) into Eq.~(\ref{eq_dEc-largemu}), which gives
\beq
\frac{\partial}{\partial \mu}\overline{E}_{c,\, {\rm SR}}^\mu[n] 
 =    -2\pi\frac{f_c(0)}{\mu^3}-4(\sqrt{2\pi}+A_1\pi)
\frac{f(0)}{\mu^4} 
+ O\left(\frac{1}{\mu^5}\right).
\label{eq_firstexpa}
\eeq
We see that to fully determine the term $\propto\mu^{-4}$ in Eq.~(\ref{eq_firstexpa})
we  have to know the value of the constant $A_1$. This constant
determines how the on-top $f^\mu(0)$ approaches the Coulombic value
$f(0)$ for large $\mu$. In fact,
from Eq.~(\ref{eq_expaf}) we have
\beq
f^\mu(0)=f(0)\left[1+\frac{1}{\mu}\left(\frac{2}{\sqrt{\pi}}+2 A_1\right)\right]
+O\left(\frac{1}{\mu^2}\right).
\label{eq_expaf0}
\eeq 
In Ref.~\onlinecite{julien}
the $\mu\to\infty$ limit of the long-range interaction was formally
rewritten as the Coulomb
interaction $1/r_{12}$ plus a perturbation~\cite{julien}
\beq
\frac{\erf(\mu r_{12})}{r_{12}}=\frac{1}{r_{12}}-\frac{\pi}{\mu^2}\delta(\rv_{12})
+O\left(\frac{1}{\mu^3}\right),
\label{eq_delta}
\eeq
where $\delta(\rv)$ is the Dirac delta function. The fact that the perturbation
is of order $\mu^{-2}$ lead to the conclusion~\cite{julien}
 that the perturbation on $\Psi^\mu$
is also of order $\mu^{-2}$ with respect to the Coulombic case, which
would correspond to
$A_1=-1/\sqrt{\pi}$ in Eq.~(\ref{eq_expaf0}). However, because
of the singular nature of the Dirac delta function, this argument does not hold
at $r_{12}=0$.
\par
\begin{figure}
\includegraphics[width=7.2cm]{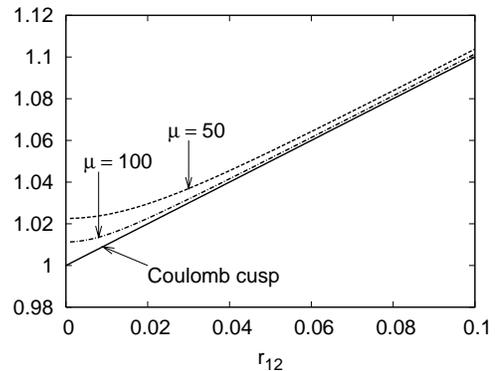} 
\caption{The Coulombic cusp $f(0)(1+r_{12})$ (with $f(0)=1$) is
compared to the expansion of $f^\mu(r_{12})$ in the $\mu\to\infty$
limit, $f^\mu(r_{12})=f(0)(1+2 r_{12} p_1(\mu r_{12})+2/(\sqrt{\pi}\mu)+
2A_1/\mu)$,
with $A_1=0$. Any value $A_1>0$ makes
$f^\mu(r_{12})$ differ from the Coulombic $f(r_{12})$ of orders $1/\mu$
also in the region $r_{12}\gg 1/\mu$. }
\label{fig_p}
\end{figure}

To determine the correction to the on-top value when $\mu\to\infty$,
here we take a large value of $\mu$ and a small value $r_{12}=r_0$ such that
$\mu^{-1}\ll r_0\ll 1$ (take, e.g. $r_0=1/\mu^{1-q}$ with $0<q<1$). 
For such value of $r_0$ we have $\mu r_0\gg 1$ so that from Eqs.~(\ref{eq_expap1}) and
(\ref{eq_expaf}) we obtain
\beq
f^\mu(r_0)=f(0)\left(1+r_0+\frac{2A_1}{\mu}
\right)+...,
\label{eq_prova}
\eeq
where the next leading terms are of order $1/\mu^2$ and/or $r_0^2$. 
We then notice that $A_1$ cannot be equal to $-1/\sqrt{\pi}$, since any
value of $A_1< 0$ would make $f^\mu(r_0)$ smaller than the
full interacting value $f(r_0)$, while,
because for small
$r_{12}$ the long-range interaction $\erf(\mu r_{12})/r_{12}$ is less repulsive than 
$1/r_{12}$, for $r_0$ small enough we expect that $f^\mu(r_0)\ge f(r_0)$.
So $A_1\ge 0$, and thus the correction to the on-top value must be of order $1/\mu$.
However, the argument of Ref.~\onlinecite{julien} should
be valid when $r_{12}\gg 1/\mu$. That is, we still expect from
Eq.~(\ref{eq_delta}) that the perturbed $\Psi^\mu$ differs from
the Coulombic $\Psi$ 
of an order higher than $1/\mu$ for $r_{12}\gg 1/\mu$.
This is achieved only if $A_1=0$, as shown by Eq.~(\ref{eq_prova}). 
 The result corresponding to $A_1=0$ 
is illustrated in Fig.~\ref{fig_p}, where we compare
the Coulomb cusp $f(0)(1+r_{12})$ to the short-range expansion of $f^\mu(r_{12})$ of
Eq.~(\ref{eq_expaf}), with $A_1=0$. Any value of $A_1$ larger
than zero makes the difference between the Coulombic $f(r_{12})$
and $f^\mu(r_{12})$ of order $1/\mu$ also in the region $r_{12}\gg 1/\mu$, while
with $A_1=0$ (as in Fig.~\ref{fig_p}) this difference is of higher order, 
as expected from Eq.~(\ref{eq_delta}).
\par
We thus conclude that
\beq
f^\mu(0)=f(0)\left(1+\frac{2}{\sqrt{\pi}\,\mu}\right)+O\left(\frac{1}{\mu^2}\right).
\label{eq_fico}
\eeq
This equation is also confirmed in the next Sec.~\ref{sec_elegas} for the
case of the high-density electron gas that can be treated exactly.
\par
\begin{figure}
\includegraphics[width=7.2cm]{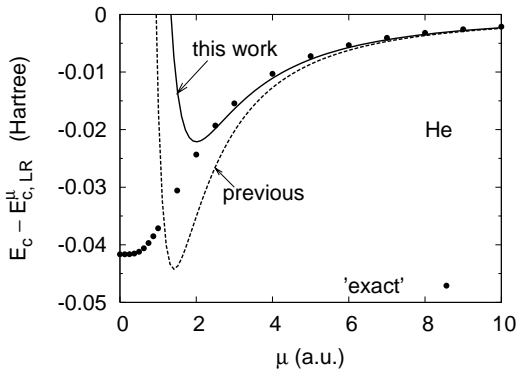} 
\caption{The ``exact'' values of $\overline{E}_{c,\, {\rm SR}}^\mu[n]=
E_c-E_{c,\, {\rm LR}}^\mu[n]$ for the He atom~\cite{sav_madrid,julien} are compared 
with the large-$\mu$ expansion of Eq.~(\ref{eq_expafinal}), and with the
previous result for the same expansion given in Eq.~(30) of Ref.~\onlinecite{julien}. 
In both expansions the ``exact'' on-top value
is taken from  Ref.~\onlinecite{cios-umri}.}
\label{fig_heexpa}
\end{figure}

The final expansion of the short-range functional $\overline{E}_{c,\, {\rm SR}}^\mu[n]$
for large $\mu$ is then
\beq
\overline{E}_{c,\, {\rm SR}}^\mu[n]=f_c(0)\frac{\pi}{\mu^2}+f(0)\frac{4\sqrt{2\pi}}{3\,
\mu^3}+O\left(\frac{1}{\mu^4}\right),
\label{eq_expafinal}
\eeq
where $f(0)$ and $f_c(0)$ are the on-top value and its correlated part,
$f_c=f-f_{\rm KS}$, of the physical system. This expansion differs
from the one of Ref.~\onlinecite{julien} by a factor $\sqrt{2}$ in the second term
(see Appendix~\ref{app_2balle} for comments on this discrepancy). 
The two expansions for the case of the He atom are compared in 
Fig.~\ref{fig_heexpa} with the ``exact'' results~\cite{sav_madrid,julien} for
$\overline{E}_{c,\, {\rm SR}}^\mu[n]=E_c-E_{c,\, {\rm LR}}^\mu[n]$. The 
``exact'' on-top value $f(0)$ is taken from Ref.~\onlinecite{cios-umri}. 
We see that the new expansion more accurately reproduce the ``exact'' data
for large $\mu$, and that the $\mu$-value for which it breaks down (i.e.,
where the expansion has its minimum) coincides with the minimum value of
$\mu$ for which it still gives accurate short-range correlation energies.
\par
In general the on-top value of the physical system, $f(0)$, is not accessible.
A plausible approximation proposed in Ref.~\onlinecite{julien} 
consists in replacing $f(0)$ in Eq.~(\ref{eq_expafinal})
with its local-density approximation (LDA) value,
\beq
f_{\rm LDA}(0)=\frac{1}{2}\int n(\rv)^2\, g(r_{12}=0;
n(\rv))\, d\rv,
\label{eq_f0LDA}
\eeq
where $g(r_{12};n)$ is the pair distribution function~\cite{GP1,GP2}
of the standard electron gas model (with Coulomb interaction $1/r_{12}$)
of uniform density $n$.
The new Eq.~(\ref{eq_fico}) allows to estimate the physical on-top value
 starting from the one of the model system $\Psi^\mu$. Potential applications
of this idea are discussed in Sec.~\ref{sec_approx}, together with simple examples.  Notice also that Eq.~(\ref{eq_fico}) is also valid locally, i.e.,
we have
\beq
\tilde{P}_2^\mu(\rv,r_{12}=0)=\tilde{P}_2(\rv,r_{12}=0)\left(1+\frac{2}{\sqrt{\pi}\,\mu}\right)+O\left(\frac{1}{\mu^2}\right),
\label{eq_ficolocale}
\eeq
where $\tilde{P}_2^\mu(\rv,r_{12})$ was defined in Eq.~(\ref{eq_p2sph}), and
$\tilde{P}_2(\rv,r_{12})$ is the pair density  of the physical system 
($\mu=\infty$).
\par 
In Appendix~\ref{app_l} we also consider the case of a fully polarized system,
for which we find that the short-range correlation energy has the
large-$\mu$ expansion
\beq
\overline{E}_{c,\, {\rm SR}}^\mu[n=n_{\uparrow}]=f_c''(0)\frac{3\pi}{8\mu^4}+f''(0)\frac{3\sqrt{2\pi}}{10\,
\mu^5}+O\left(\frac{1}{\mu^6}\right),
\label{eq_expafinalz1}
\eeq
where $f''(0)$ and $f_c''(0)$ are the second derivative at
$r_{12}=0$ and its correlated part of
the physical, fully interacting, system~\cite{Yaspar}. We also found that, again only
in the case of a fully polarized system, the second derivative of 
$f^\mu(r_{12})$ at $r_{12}=0$
approaches the one of the Coulombic system as
\beq
(f^\mu)''(0)=f''(0)\left(1+\frac{2}{3\sqrt{\pi}\,\mu}\right)+O\left(\frac{1}
{\mu^2}\right).
\label{eq_expacurv}
\eeq

\section{On-top pair density of a uniform system 
with interaction $\erf(\mu r_{12})/r_{12}$  }
\label{sec_elegas}
Before coming to applications, we consider the special case
of the uniform electron gas, for which something more than
Eq.~(\ref{eq_fico}) can be done. 
We consider a uniform system with long-range-only
interaction,
\begin{equation}
H^{\mu}  =  -\frac{1}{2}\sum_{i=1}^N \nabla^2_{\rv_i}+V_{\rm LR}^\mu
+V_{eb}^{\mu}+V_{bb}^{\mu},
\label{eq_ham}
\end{equation}
where $V_{\rm LR}^\mu$ is the modified electron-electron interaction
\begin{equation}
V_{\rm LR}^\mu  =  \frac{1}{2}\sum_{i\ne j=1}^N\frac{\erf(\mu|\rv_i-\rv_j|)}
{|\rv_i-\rv_j|}, 
\label{eq_Vee}
\end{equation}
$V_{eb}^{\mu}$ is, accordingly, 
the interaction between the electrons and a rigid,
positive, uniform background of density $n=(4\pi r_s^3/3)^{-1}$
\begin{equation}
V_{eb}^{\mu}  =  -n\sum_{i=1}^N\int d\xv \,\frac{\erf(\mu|\rv_i-\xv|)}
{|\rv_i-\xv|}, 
\label{eq_Veb}
\end{equation}
and $V_{bb}^{\mu}$ is the corresponding background-background interaction
\begin{equation}
V_{bb}^{\mu}  =  
\frac{n^2}{2}\int d\xv\int d\xv' \,\frac{\erf(\mu|\xv-\xv'|)}
{|\xv-\xv'|}.
\label{eq_Vbb}
\end{equation}
When $\mu\to\infty$ $H^{\mu}$ tends to the standard
jellium hamiltonian, while when $\mu\to 0$ we recover the
noninteracting electron gas.

We focus  on the $\mu$-dependence of the on-top value of the pair-distribution 
function~\cite{GP1,GP2}
$g(r_{12}=0,r_s,\mu)$,
which has its own interest to construct the LDA
approximation for the long-range and short-range functionals, and
for spin-density functional theory in the framework of the
alternative on-top interpretation~\cite{ontop}. The relation between
the function $g$ and
the function $f$ of Eq.~(\ref{eq_intra}) is $g=2 f/n N$.
\begin{figure}
\includegraphics[width=7.2cm]{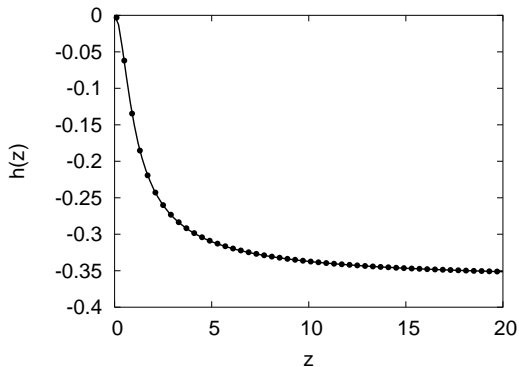} 
\caption{The function $h(z)$ that determines the 
high-density limit of the on-top pair density of the ``erf'' 
gas [see Eq.~(\ref{eq_h})].
The numerical evaluation of Eq.~(\ref{eq_gc0HD}) (points) is compared
to the fitting function of Eq.~(\ref{eq_padeh}) (solid line).}
\label{fig_h}
\end{figure}
\subsection{High-density limit}
As in the Coulomb gas, by switching to scaled
units $\sv_i=\rv_i/r_s$, we see that when
$r_s\to 0$ the potential of Eqs.~(\ref{eq_Vee})-(\ref{eq_Vbb})
becomes a perturbation to the noninteracting gas.
Defining the correlation
contribution to the on-top value,
$g_c(0,r_s,\mu)=g(0,r_s,\mu)-\frac{1}{2}$,  and
following Kimball~\cite{kimbHD}, the first-order correction
(with respect to the interaction potential) to the on-top
pair density is
\beq
g_c(0,r_s\to 0,\mu)=6 \int_0^{\infty} \Delta S_D(t,\mu)\, t^2\, dt +...,
\label{eq_gc0HD}
\eeq
where $\Delta S_D(t,\mu)$ is the direct second-order contribution
to the static structure factor~\cite{kimbHD},
\begin{eqnarray}
&  \Delta S_D(q,\mu)= \nonumber \\
& \frac{4}{N}\sum_{\kv\,\sigma,\kv'\,\sigma'}
v_{ee}(q) \frac{n_{\rm F}(\kv)n_{\rm F}(\kv')[1-n_{\rm F}(\kv+\qv)][1-n_{\rm F}(\kv'-\qv)]}
{k^2+k'^2-(\kv+\qv)^2-(\kv'-\qv)^2},
\label{eq_SD}
\end{eqnarray}
$n_{\rm F}$ is the usual Fermi occupation function~\cite{kimbHD} and
$v_{ee}(q)$ is the Fourier transform of the electron-electron interaction.
In Eq.~(\ref{eq_gc0HD}) the scaled variable $t=q/2k_F$ 
[$k_F=(\alpha r_s)^{-1}$, $\alpha=(\frac{4}{9\pi})^{1/3}$] has been used.
The function $\Delta S_D(t,\mu)$ is thus equal to the one
computed by Kimball~\cite{kimbHD} and reported in his Eq.~(11), except
for a multiplying factor $e^{-t^2 k_F^2/\mu^2}$ coming from
the Fourier transform of $\erf(\mu r_{12})/r_{12}$. From Eqs.~(\ref{eq_gc0HD})
and (\ref{eq_SD}) we find
\beq
g_c(0,r_s\to 0,\mu)= r_s \,h(\mu/k_F)+...,
\label{eq_h}
\eeq
where
the function $h(z)$ has the following asymptotic behaviors
\begin{eqnarray}
h(z\to 0) & = & -\tfrac{6\alpha}{\pi} (1-\ln 2)\,z^2+O(z^3) \\
h(z\to \infty) & = & a_{\rm HD}+\frac{\alpha}{\sqrt{\pi}\,z}+O(z^{-2}),
\label{eq_g0HDlargemu}
\end{eqnarray}
and
$a_{\rm HD}=-\alpha (\pi^2 + 6\ln 2-3)/5\pi\approx -0.36583$
is the high-density limit of the standard jellium model.
Notice that Eq.~(\ref{eq_g0HDlargemu}) confirms, for the high-density
electron gas, Eq.~(\ref{eq_fico}).
\par
For intermediate values of $z$ 
we computed numerically the function $h(z)$, and we found
that it can be accurately fitted by the Pad\'e form
\beq
h(z)=\frac{a_1\, z^2+a_2\, z^3}{1+b_1\,z+b_2\,z^2+b_3\,z^3},
\label{eq_padeh}
\eeq
with $a_1= -\tfrac{6\alpha}{\pi} (1-\ln 2)$, $b_1=1.4919$,
$b_3=1.91528$, $a_2=a_{\rm HD}\,b_3$, $b_2=(a_1-b_3 \alpha/\sqrt{\pi})/a_{\rm HD}$. 
The numerical results and the fitting function of Eq.~(\ref{eq_padeh})
are reported in Fig.~\ref{fig_h}.
 
\begin{figure}
\includegraphics[width=7.2cm]{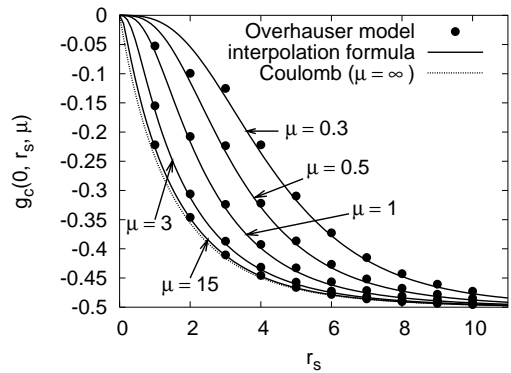} 
\caption{The correlation on-top pair density of the electron
gas with interaction $\erf(\mu r_{12})/r_{12}$ for different $\mu$ 
(in a.u.) as a function
of the dimensionless density parameter $r_s$. The results from the Overhauser model
are compared with the interpolation formula of Eq.~(\ref{eq_formula}). The
dotted line corresponds to the standard jellium model with interaction
$1/r_{12}$.}
\label{fig_ontop}
\end{figure}

\subsection{Interpolation formula}
The high-density limit of Eq.~(\ref{eq_h}) and of
Fig.~\ref{fig_h} tells us how (at least for small $r_s$)
$g_c(0,r_s,\mu)$ approaches 
the two limits, the noninteracting gas ($\mu\to 0$)
and the Coulomb gas ($\mu\to\infty$).

A simple interpolation formula for all densities can be
built by assuming that the $\mu$ dependence of $g_c(0,r_s,\mu)$
is roughly the same at each $r_s$.  
We thus start from the parameterization of the on-top pair
density of the jellium model given in Ref.~\onlinecite{GP1},
\beq
g(0,r_s,\mu=\infty)=\frac{1}{2}(1-B r_s+Cr_s^2+Dr_s^3+Er_s^4)\,e^{-dr_s},
\eeq
where  $C=0.08193$, $D=-0.01277$,
$E=0.001859$, $d=0.7524$ and $B=-2 \,a_{HD}-d$,
and we simply rescale homogeneously all the coefficients with the function
$h(z=\mu/k_F)/a_{\rm HD}$,
\begin{eqnarray}
g_c(0,r_s,\mu) & = & \frac{\,
e^{-d\,r_s\,h(z)/a_{\rm HD}}}{2}\Big[1- B\, \frac{h(z)}{a_{\rm HD}} r_s+
 C \,\frac{h(z)^2}{a_{\rm HD}^2} r_s^2
\nonumber \\ 
& & + D\,\frac{h(z)^3}{a_{\rm HD}^3} r_s^3
+ E \,\frac{h(z)^4}{a_{\rm HD}^4} r_s^4\Big]-\frac{1}{2}.
\label{eq_formula}
\end{eqnarray}
This simple guess smoothly interpolates between the $\mu\to 0$ and
$\mu\to\infty$ limits, and is exact when $r_s\to 0$. 

\subsection{Results from the Overhauser model}
To check the validity of the interpolation formula of
Eq.~(\ref{eq_formula}) we evaluated the on-top
$g_c(0,r_s,\mu)$ within the 
``extended Overhauser model''~\cite{Ov,GP1},
which gave good results for the
standard jellium model~\cite{GP1} and for
two-electron atoms~\cite{GS1}. Notice that the on-top value is not
known exactly. The differences between
the jellium on-top pair densities from different approximate methods
(including the ``extended Overhauser model'')
are discussed in Refs.~\onlinecite{pisaniontop,cios_ladd,qian}. 

The scattering equations of the ``extended Overhauser model''
are widely explained in Refs.~\onlinecite{GP1,DPAT}. Here we
simply solved the same equations with the electron-electron
interaction $\erf(\mu r_{12})/r_{12}$ screened by a sphere of radius
$r_s$ of uniform positive charge density $n$ and attracting
the electrons with the same modified interaction,
\beq
V_{\rm eff}(r_{12},r_s,\mu)=\frac{\erf(\mu r_{12})}{r_{12}}
-
\int_{|\rv'|\le r_s} 
n\,  \frac{\erf(\mu|\rv' - \rv_{12}|)}{|\rv' - \rv_{12}|}\,d \rv'.
\label{eq_veff}
\eeq
This potential is reported in the Appendix of Ref.~\onlinecite{GS1},
where it has been used for two-electron
atoms with rather accurate results for the corresponding short-range
correlation energy. $V_{\rm eff}(r_{12},r_s,\mu)$ is a screened potential that
tends to the ``Overhauser potential''\cite{Ov,GP1} when $\mu\to\infty$,
and which goes to zero when $\mu\to 0$. As in the
original Overhauser model, the idea behind Eq.~(\ref{eq_veff}) 
is that the radius of the screening ``hole'' is exactly equal to $r_s$.

The results for the on-top $g_c(0,r_s,\mu)$ from the Overhauser
model are reported in Fig.~\ref{fig_ontop} as a function of
$r_s$ for different values of $\mu$. We see that the the simple
interpolation formula of Eq.~(\ref{eq_formula}) accurately
captures the $\mu$ and $r_s$ dependence of $g_c(0,r_s,\mu)$.

\section{Applications, perspectives, and conclusions}
\label{sec_approx}
The main results of this work are (i) the corrected expansion of the short-range
correlation energy functional, Eq.~(\ref{eq_expafinal}), (ii) the expansion of the
on-top pair density of Eqs.~(\ref{eq_fico})
and (\ref{eq_ficolocale}), and (iii) the parameterization of  
the $\mu$- dependence of the on-top pair density of the 
uniform electron gas, Eq.~(\ref{eq_formula}).
All these results (i-iii) 
can be useful for the construction of approximate short-range
correlation energy functionals: \par
\begin{figure}
\includegraphics[width=7.2cm]{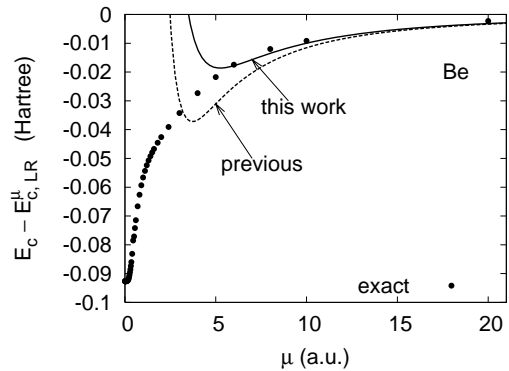} 
\caption{The ``exact'' values of $\overline{E}_{c,\, {\rm SR}}^\mu[n]=
E_c-E_{c,\, {\rm LR}}^\mu[n]$ for the Be atom~\cite{sav_madrid,julien} are compared 
with the large-$\mu$ expansion of Eq.~(\ref{eq_expafinal}), and with the
previous result for the same expansion given in Eq.~(30) of Ref.~\onlinecite{julien}. 
In both expansions the ``exact'' on-top value
is taken from  Ref.~\onlinecite{cios-umri}.}
\label{fig_Beexpa}
\end{figure}

(i) In Ref.~\cite{julien}
the large-$\mu$ expansion of the correlation energy functional has been already
used to construct approximations: the idea is~\cite{julien} 
to interpolate between a given density functional approximation (DFA) of
standard KS theory~\cite{ladder} at $\mu=0$, and the $\mu\to\infty$ 
expansion of
$\overline{E}_{c,\, {\rm SR}}^\mu[n]$. In the spirit of the usual DFT 
approximations, this interpolation is done locally, i.e.,
\beq
\overline{E}_{c,\, {\rm SR}}^\mu[n]=\int d\rv \,n(\rv)\,
\overline{\epsilon}_{c,\,\rm SR}^\mu(\rv),
\label{eq_eclocint}
\eeq 
where $\overline{\epsilon}_{c,\,\rm SR}^\mu(\rv)$ is built, e.g., 
as~\cite{julien}
\beq
\overline{\epsilon}_{c,\,\rm SR}^\mu\approx\frac{\ec^{\rm DFA}}
{1+d_1\,\mu+d_2\,\mu^2}.
\label{eq_interpolazione}
\eeq
The parameters $d_1$ and $d_2$ are fixed by the condition that 
Eq.~(\ref{eq_eclocint}) recovers the correct $\mu\to\infty$ expansion
of $\overline{E}_{c,\, {\rm SR}}^\mu[n]$, and $\ec^{\rm DFA}$ can be, e.g.,
the PBE correlation functional~\cite{PBE} of standard Kohn-Sham theory or
any other available approximation. This correlation functional can be
combined with a similar interpolation for exchange~\cite{julien}, or
with the exchange functional of Heyd, Scuseria, and 
Ernzerhof~\cite{scuseria}. This way of constructing approximations
should be improved by using the corrected
expansion of  Eq.~(\ref{eq_expafinal}), as suggested by Fig.~\ref{fig_heexpa}.
In Fig.~\ref{fig_Beexpa} we also show similar data for the Be atom: again,
the corrected expansion is closer to the ``exact'' 
data~\cite{sav_madrid,julien} at large $\mu$ than the previous expansion
used in Ref.~\onlinecite{julien}. 
\par 
(ii) To impose the correct large-$\mu$ expansion in approximations like
the one of
Eq.~(\ref{eq_interpolazione}) we need an estimate of the physical 
($\mu=\infty$) on-top
pair density $\tilde{P}_2(\rv,r_{12}=0)$. In Ref.~\onlinecite{julien}
the LDA approximation 
[the integrand of the right-hand side of Eq.~(\ref{eq_f0LDA})] was used.
The new Eq.~(\ref{eq_ficolocale}) allows to use the on-top pair density of 
the partially correlated
wavefunction $\Psi^\mu$ to estimate $\tilde{P}_2(\rv,r_{12}=0)$. In fact, once
we have made a calculation at a given (moderately large) $\mu$ (say, $\mu=\mu_0$) we
can estimate $\tilde{P}_2(\rv,r_{12}=0)$ as
\beq
\tilde{P}_2(\rv,r_{12}=0) \approx \tilde{P}_2^{\mu_0}(\rv,r_{12}=0)
\left(1+\frac{2}{\sqrt{\pi}\mu_0}\right)^{-1}.
\label{eq_approxf0}
\eeq
 There are cases in which this estimate could be much better
than the one obtained by using the LDA approximation for the physical on-top
pair density. In fact, the use of the partially correlated 
$\tilde{P}_2^\mu(\rv,r_{12}=0)$ would
correct the self-interaction error of LDA, becoming exactly equal to zero
for any one-electron density.
Consider the example of the stretched H$_2$ molecule, for which 
the estimate from Eq.~(\ref{eq_approxf0}) would be essentially exact (equal to zero for any $\mu>0$),
 while LDA gives a spurious nonzero on-top value, unless we consider the
spin broken-symmetry solution. 

To show that Eq.~(\ref{eq_approxf0}) gives indeed a quantitative
 reliable estimate of the physical on-top pair density we have considered the 
simple example of the He atom, and we have inserted Eq.~(\ref{eq_approxf0})
directly in the expansion of Eq.~(\ref{eq_expafinal}):
 the error on the estimated short-range correlation energy
at $\mu_0=2.5$ is 3~mH;
if we choose $\mu_0=2$ the error in   $\overline{E}_{c,\, {\rm SR}}^{\mu_0}[n]$ is 5~mH, and for  $\mu_0=1$ is 11~mH. Of course, when $\mu_0$ becomes too small, the
large-$\mu$ expansion of Eqs.~(\ref{eq_expafinal}) and 
(\ref{eq_approxf0}) is no longer valid. 
\par
(iii) The on-top pair density of the uniform electron gas with long-range-
only interaction of Eq.~(\ref{eq_formula}) can be used, in combination with 
the correlation energy of the spin-polarized long-range electron gas~\cite{simone}, to
implement the local approximation for the 
on-top pair density interpretation of spin density functional theory~\cite{ontop}. 
\par
Another interesting application of Eq.~(\ref{eq_formula}) is connected to point (ii): 
the construction of functionals that explicitly
depend on the on-top $f^\mu(0)$ [or locally on $\tilde{P}_2^\mu(\rv,r_{12}=0)$]
of the partially correlated wavefunction could
use the $\mu$ dependence of the on-top LDA value to go beyond Eq.~(\ref{eq_approxf0}),
\beq
f_c(0)\approx f^{\mu_0}_c(0) \frac{\int n(\rv)^2\, g_c(0,
r_s(\rv),\mu=\infty)\, d\rv}{\int n(\rv)^2\, g_c(0,
r_s(\rv),\mu_0)\, d\rv}.
\label{eq_approxLDA}
\eeq
For example for the He atom Eq.~(\ref{eq_approxLDA}) at $\mu_0=2$ gives 
$f_c(0)=-0.086$, while Eq.~(\ref{eq_approxf0}) gives $-0.090$. The
corresponding ``exact'' value~\cite{cios-umri} is $-0.085$.
Local versions of Eq.~(\ref{eq_approxLDA}) can be also considered, e.g.,
\beq
f_c(0)\approx  \int n(\rv)^2\, \tilde{P}_{2,c}^{\mu_0}(\rv,r_{12}=0) \frac{g_c(0,
r_s(\rv),\mu=\infty)}{g_c(0,
r_s(\rv),\mu_0)}\, d\rv,
\label{eq_approxlocale}
\eeq
where $\tilde{P}_{2,c}^{\mu_0}(\rv,r_{12})$ is obtained
by subtracting the Kohn-Sham pair density from  $\tilde{P}_2^{\mu_0}$.
Again, the advantage of including
in the construction of short-range functionals the on-top
$\tilde{P}_{2,c}^{\mu}(\rv,r_{12}=0)$ is to locally remove the 
self-interaction error.
\par
In conclusions, we have presented a comprehensive study of the short-range
behavior of systems interacting with the potential 
$\erf(\mu r_{12})/r_{12}$, in
connection with the properties of long- and short-range correlation 
energy density
functionals. The same kind of analysis can be of course repeated for other
splittings of the Coulomb electron-electron interaction~\cite{vaffa,erfgau}.
Future work will be mainly oriented to the exploration of the promising
approach of short-range functionals that explictly depend
on $\tilde{P}_2^\mu(\rv,r_{12}=0)$. 

\section*{Acknowledgments}
We thank J. Toulouse for valuable discussions and for the data of 
Ref.~\onlinecite{sav_madrid}.

\appendix
\section{Alternative derivation of  Eq.~(\ref{eq_expaf})}
\label{app_2balle}
Start from Eq.~(\ref{eq_relative}), and consider the following series 
expansions around $x=0$
\begin{eqnarray}
u^\mu(x) & = & \sum_{n=0}^{\infty} a_n(\mu) x^{n+1},
\label{eq_user} \\
\frac{\erf(\mu x)}{x} & = & \sum_{n=0}^\infty b_n \mu (\mu x)^{2 n},
\nonumber \\
b_n & = & \frac{2}{\sqrt{\pi}}\frac{(-1)^n}{(2n+1)n!}.
\label{eq_erfser}
\end{eqnarray}
This last series has an infinite radius of convergence for any finite $\mu$.
\par
The complicated operator
$\mathcal{E}^\mu$ can be also expanded in powers of $x$ around $x=0$. Its
expansion will only contain even powers of $x$ because the hamiltonian of 
Eq.~(\ref{eq_Hmu}) is even in ${\bf x}=\rv_{12}$. As expected, 
the expansion
of  $\mathcal{E}^\mu$ does not
play any role, so we do not consider its non-spherical components (moreover,
in the end we are only interested in the spherically averaged pair density). The only
important requirement is that $\mathcal{E}^\mu$ remains bounded when
$\mu\to\infty$ and $x\to 0$, as it happens for the Coulomb 
interaction~\cite{kato,kimball}. 
We thus write
\beq
\mathcal{E}^\mu=\sum_{k=0}^\infty e_{2k}(\mu) x^{2k}.
\label{eq_emuser}
\eeq
By inserting Eqs.~(\ref{eq_user}), (\ref{eq_erfser}) and~(\ref{eq_emuser}) 
into Eq.~(\ref{eq_relative}) we find that the
$a_n(\mu)$ with odd $n$ are zero (as expected from the fact that 
$\erf(\mu x)/x$ is an even function of $x$), while the even $n$ coefficients
with $n\ge 2$ diverge as $\mu$ increases, and, to leading
order when $\mu\to\infty$, they are all proportional to $a_0(\mu)$,
\begin{eqnarray}
a_{2k+2}(\mu) & = & a_0(\mu)\Biggr[\frac{b_k\mu^{2k+1}}{(2k+2)(2k+3)}+
\frac{\mu^{2k}}{(2k+2)(2k+3)} \nonumber \\
& \times & \sum_{i=1}^k\frac{b_{k-i}b_{i-1}}{2i(2i+1)}\Biggl]
+O(\mu^{2k-1}).
\label{eq_recan}
\end{eqnarray}
This relation shows that $\psi^\mu(x)$ has the structure
\beq
\psi^\mu(x)=a_0(\mu)[1+x\,p_1(\mu \,x)+x^2\,p_2(\mu\,x)+...]
\eeq
where
\begin{eqnarray}
p_1(y) & = & \sum_{k=0}^{\infty}
\frac{b_k y^{2k+1}}{(2k+2)(2k+3)} \label{eq_altp1} \\
p_2(y) & = & \sum_{k=1}^\infty\left(
\frac{y^{2k}}{(2k+2)(2k+3)}\sum_{i=1}^k\frac{b_{k-i}b_{i-1}}{2i(2i+1)}
\right).
\end{eqnarray}
Equation~(\ref{eq_altp1}) gives exactly the same function $p_1(y)$ of
Eq.~(\ref{eq_p1}), and Eq.~(\ref{eq_recan}) confirms that all the terms beyond
the ones considered in Eq.~(\ref{eq_expaf}) contribute to Eq.~(\ref{eq_firstexpa})
to orders $\mu^{-5}$ or higher.
\par
We can now clearly see where the discrepancy with the result of 
Ref.~\onlinecite{julien} comes from: the small-$r_{12}$ expansion of $f^\mu(r_{12})$
only contains even powers of $r_{12}$ for any finite $\mu$. It is thus incorrect
to insert the odd coefficient $c_1$ of the Coulombic system in Eq.~(\ref{eq_dEc-largemu}), as it was done in Ref.~\onlinecite{julien}. What happens, instead, is that 
all the even coefficients of the small-$r_{12}$ expansion of  $f^\mu(r_{12})$
diverge for large $\mu$ and they all contribute to the term $\propto \mu^{-4}$
in Eq.~(\ref{eq_dEc-largemu}).
Furthermore, in Ref.~\onlinecite{julien} it was assumed that the leading
order in the large-$\mu$
expansion of the on-top value $f^\mu(0)$ is $1/\mu^2$, while we have shown that
the correction to $f^\mu(0)$ with respect to the Coulombic case is of order
$1/\mu$.

\section{The case $\ell =1$}
\label{app_l}
We insert the expansion of $u^\mu(y)$ for large $\mu$ of Eq.~(\ref{eq_expau}) into
the equivalent of Eq.~(\ref{eq_relativey}) for the case $\ell=1$,
\beq
\left[-\frac{d^2}{dy^2}+\frac{2}{y^2}+
\frac{1}{\mu}\frac{\erf(y)}{y}\right]u^\mu(y)=\frac{1}{\mu^2}\mathcal{E}^\mu u^\mu(y).
\label{eq_relativeyl1}
\eeq
The condition that
the left-hand side be of order $1/\mu^2$ yields
\begin{eqnarray}
u^{(\infty)}(y) & = & b\,y^2 \\
\left[\frac{d^2}{dy^2}-\frac{2}{y^2}\right]u^{(-1)}(y) & = &
b\,y\,\erf(y).
\label{eq_u1trip}
\end{eqnarray}
By solving Eq.~(\ref{eq_u1trip}) we find that the intracule 
$f^\mu(r_{12})$ has, for large $\mu$,  the small-$r_{12}$ expansion
\beq
f^\mu(r_{12})=\frac{f''(0)}{2}\,r_{12}^2\,\left[1+2r_{12}q_1(\mu r_{12})+
\frac{2}{3\sqrt{\pi}\mu}+\frac{B_1}{\mu}\right],
\eeq
where the function $q_1(y)$ is equal to
\begin{eqnarray}
q_1(y)& = & \frac{e^{-y^2}(2y^2-1)}{8\sqrt{\pi} y^3}-\frac{1}{3\sqrt{\pi}\,y}+
\frac{\erf(y)(4y^4+1)}{16 y^4},
\label{eq_q1}
\end{eqnarray}
and $B_1$ is a constant of integration that is not determined by the
requirement that $f^\mu$ vanishes at $r_{12}=0$.
The function $q_1(y)$ has the asymptotic behaviors
\begin{eqnarray}
q_1(y\to 0) & = & \frac{y}{5\sqrt{\pi}}+O(y^3) \\
q_1(y\to\infty) & = & \frac{1}{4}-\frac{1}{3\sqrt{\pi}\,y}+O\left(\frac{1}{y^3}\right).
\end{eqnarray}
Again, we see that if we fix $r_{12}=r_0\ll 1$, and then let $\mu\to\infty$
we find $f^\mu(r_0)\propto r_0^2[1+r_0/2+...]$, which is the parallel-spin
cusp condition for the Coulomb interaction~\cite{kimball}. 
But for any finite $\mu$
we have, for small $r_{12}$,  $f^\mu(r_{12})\propto r_{12}^2[1+r_{12}^2 2\mu/5\sqrt{\pi}+...]$. The proof that $B_1=0$, and thus of Eqs.~(\ref{eq_expafinalz1}) 
and~(\ref{eq_expacurv}) is then completely
analogous to the one for the case $\ell=0$.

\end{document}